# Sub-micron strain analysis of local stick-slip motion of individual shear bands in a bulk metallic glass

I. Binkowski[1], S. Schlottbom[1], J. Leuthold[1], S. Ostendorp[1], S.V. Divinski[1], G. Wilde[1, 2]

[1] Institute of Materials Physics, University of Münster, Wilhelm-Klemm-Str. 10, 48149 Münster, Germany

[2] Institute of Nanochemistry and Nanobiology, Shanghai University, Shanghai 200072, China

Nanodot deposition on a side surface of a rectangular sample and digital image correlation are used to quantify the in-plane strain fields associated with the propagation of a shear band in a PdNiP bulk metallic glass, induced by rolling. Within the resolution of the method related to an average inter-dot distance of 100 nm, deformation is found to be highly localized at the shear bands, while alternating areas with a size of 100 - 400 nm with opposite local shear strains are found. This phenomenon substantiates a local stick-slip nature of shear band propagation during the metallic glass deformation, even during rolling.

Metallic glasses deform inhomogeneously via the formation and activation of thin shear bands (SBs) if plastic strain is applied at low homologous temperatures $T_h < 0.7$, where $T_h$ is defined with respect to the glass transition temperature $T_g$ as $T_h = T/T_g$ [1,2]. The shear bands represent regions of shear softening [3] and drastically enhanced atomic diffusivity [4] with thicknesses of the order of 10 nm [5–7] and lead to catastrophic failure via crack formation and propagation along the SBs upon continued unconstrained straining [8]. When deformed uniaxially in tension or compression at sufficiently high $T_h$, plastic deformation along the shear bands has been shown to proceed in a stick-slip manner, i.e. the slip is not proceeding in a steady-state but is alternatingly arrested and re-activated [9]. It has also been shown that this behavior is also found during the deformation of a massive metallic glass specimen that included only a single shear band [9]. In the prevalent view of shear band activation, as also summarized in recent reviews [8,10–12], the stick-slip behavior is seen as the arrest or activation of the entire shear band, i.e. the entire material inside a given shear band either is arrested or participates in the slip that yields the macroscopically observed stress drop. Additionally, it has been proposed that the presence of a stress gradient is a pre-requisite for the stick-slip mode of deformation to occur [13].

However, recent investigations by correlative analytical transmission electron microscopy has revealed alternating regions of bright or dark contrast in a single shear band that had been densified or dilated during slip induced by cold rolling at room temperature [6,7]. These regions of different contrast also showed alternating positive or negative deflections form the main propagation direction of the SB, which are of the order of about 5°. Similar contrast changes in single shear bands have already been reported in literature, yet without analyzing the reason for the contrast alterations [2] and in fact with prime focus on the dilated areas, in which even nanometer void formation has been reported [14]. The complex structure of individual shear bands has been discussed in agreement with the relaxation behavior of deformed glasses as analyzed by measuring the Boson peak contribution to the specific heat at low temperatures [15,16] and by annealing-time dependent measurements of the shear band diffusivity that have been carried out by radiotracer diffusion analyses [17].

The results of these analyses, and specifically the observation of alternating regions along a shear band that are densified or dilated with respect to the glass matrix, suggest that slip proceeded inhomogeneously, even along this single shear band and also under rolling conditions as the deformation pathway, i.e. under deformation conditions that, according to the literature [18,19], might not allow for stick-slip behavior to occur.

In an analogy to granular media, alternating densification or dilatation of initially homogeneous media would suggest that the slip velocity varied along the shear band [20,21], similar to a localized stick-slip propagation mode of individual segments of a given shear band, in contrast to the accepted view of shear band propagation in a bulk metallic glass. This apparent contradiction presents the starting point of the current analysis.

In order to analyze the local strain response of individual shear bands, bulk samples of a $Pd_{40}Ni_{40}P_{20}$ glass that, due to its high kinetic stability [22] is stable against nanocrystal formation upon deformation [23], has been deformed to different total strains by rolling at room temperature and the in-plane components of the strain tensor parallel to the surface of the specimens were investigated using digital image correlation (DIC) [24] at sub-micron spatial resolution. The results clearly indicate the presence of alternating regions along individual shear bands that are in different strain states, confirming the inhomogeneous propagation of individual shear bands and also verifying the presence of stick-slip type propagation under rolling conditions.

A Pd-Ni-P-based bulk metallic glass sample with the composition of $Pd_{40}Ni_{40}P_{20}$ (in at.%) was prepared by direct melting of palladium (purity 99.95 %) and $Ni_2P$ powder (purity 99.5 %) in an alumina crucible using an induction furnace in a purified argon atmosphere. The chemical composition of the alloy sample wasconfirmed by atomic absorption spectroscopy (Mikroanalytisches Labor Pascher, Germany). The crystalline master alloy was re-melted and then chill-cast into a copper mold with a 1x10x30 $mm^3$ cavity. X-ray diffraction was performed using a Siemens D-5000 diffractometer equipped with a Cu cathode, a $K_\alpha$ monochromator, a rotating sample holderin the θ-2θ geometry and a point detector. The as-prepared sample was checked to be fully X-ray amorphous and the plastic deformation did not result in any detectablecrystallization according to the X-ray study. Earlier investigations by TEM and also the observation of an increased low-temperature excess heat capacity (Boson peak) upon plastic deformation have also confirmed the stability of the amorphous state of this alloy against deformation-induced crystal formation [16]. Characteristic changes of the calorimetric signal upon plastic deformation were intensively analyzed elsewhere [25] and a high stability of the PdNiP glass against nanocrystallization was substantiated.

In order to analyze the components of the in-plane strain tensor locally, alumina masks consisting of regular pore arrays with a pore diameter of 50 nm and an inter-pore distance of 100 nm have been fabricated by anodic oxidation of polished high purity Al foils in an electrochemical set-up with oxalic acid as electrolyte. The porous mask was then under-etched and has been transferred onto the polished side surface of a rectangular bulk metallic glass sample in a lift-off process to serve as deposition mask. The mask, due to its small thickness of approximately 150 nm is thoroughly fixed by Van der Waals forces. Figure 1a shows a scanning electron microscopy (FEI NanoSEM 230; SEM) overview of the deposition mask. A detailed description of the mask preparation process has been published elsewhere [26]. After transfer of the mask onto the surface perpendicular to the transverse direction of the specimen (see schematic illustration in Fig. 1b), Au has been deposited by electron-beam induced evaporation (Edwards Fl 400) through the pore channels of the mask after depositing a thin layer of Ti (thickness of about 4 nm) first, to obtain a regular pattern of Au nanodots on the surface normal to the transverse direction (TD, Fig. 1b) of the specimens. After the deposition process, the mask has been mechanically removed by blowing dry air onto the surface and the pattern of the Au nanodots has been captured by SEM (Fig. 1c).

A set of glassy samples with applied Au nanodot patterns was plastically deformed by cold rolling in a two-high rolling mill in one step at room temperature to various strains. The deformation degree ε was determined by the thickness reduction and the strain rate $\dot{\varepsilon}$ was estimated to be about 5.5 s$^{-1}$. Cold rolling of the glassy samples led to shear band formation, which was directly detected by optical microscopy of the shear offset on the surfaces of the specimen. Plastic deformation to a relatively low strain of about 8 % was applied to avoid massive corrugation of the surfaces and the resulting shear band density, $\rho_{SB}$, was estimated to be about $\rho_{SB}$= 0.06 µm$^{-1}$. Note that this is the strain used in experiments on diffusion enhancement of atomic transport along the shear bands in [17]. After the deformation, the samples were again examined by SEM. Corresponding frames of SEM images recorded at the same area of the sample before and after deformation are shown in Fig. 2a and b, respectively. The blue stars represent the centers of automatically identified nanodots and a high quality of the image processing by DIC is seen.

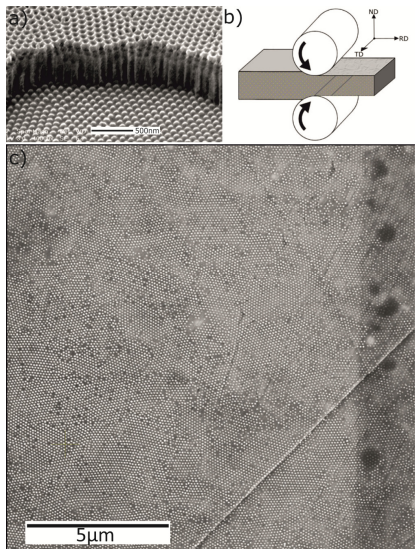

Figure 1: a) SEM image of an aluminum mask with a pore diameter of 50 nm and an inter-pore distance of 100 nm with Au deposited through the pore channels to obtain a regular pattern of Au nanodots on the surface. b) Scheme of the cold-rolling process with nanodot pattern visible on the surface normal to transverse direction . c) SEM frame of the side surface of the sample after cold rolling. Besides the nanodot pattern, shear bands are clearly visible.

Clearly, the nanodot pattern has been retained after the deformation. A regularity analysis based on a Voronoi construction coupled to Delauny triangulation [27] was applied to the dot patterns, which served to ensure identifying the same areas before and after straining to allow correlating the positions of the Au nanodots before and after plastic deformation by using digital image correlation and which also enhances the alterations of the pattern due to localized strain. For performing digital image correlation, a Matlab$^{TM}$-based code was developed that automatically identified the Au nanodots and calculated the geometric center of each dot. With the geometric arrangement of the dots and particularly with geometric imperfections such as missing dots, the algorithm identifies the translational and/or rotational shifts between the two images taken before and after the deformation, since small shifts of the position at which the SEM image is taken cannot be avoided.

A shear band can easily be identified along the line from the top-right to the bottom-left corners in the image as well as an appearance of characteristic steps on the surface (scratches), which were left on purpose after polishing, Fig. 2b. One recognizes immediately that shear displacements along the particular SB reach about 100 nm at the two particular locations.

To ensure that the measurements of the displacements are not significantly affected by artefacts stemming from the re-insertion of the specimen into the SEM (e.g. by deviations of the tilt or the imaging conditions), the undeformed specimen has been repeatedly taken out of and re-introduced into the SEM chamber without any intermediate treatment. Each time, an image of the nanodot pattern was taken and subsequently DIC was performed. The results of this analysis allowed estimating the error margin of the method that amounts to about ±0.03 (or ±3%), i.e. apparent strains of that magnitude are introduced into the analysis by deviations of the imaging conditions during imaging the sample's surface before and after deformation. As shown below, the measurements and analyses of the strain components near the shear bands exceed this confidence limit by far and thus, this systematic error does not influence any of the conclusions drawn from the present experiments.

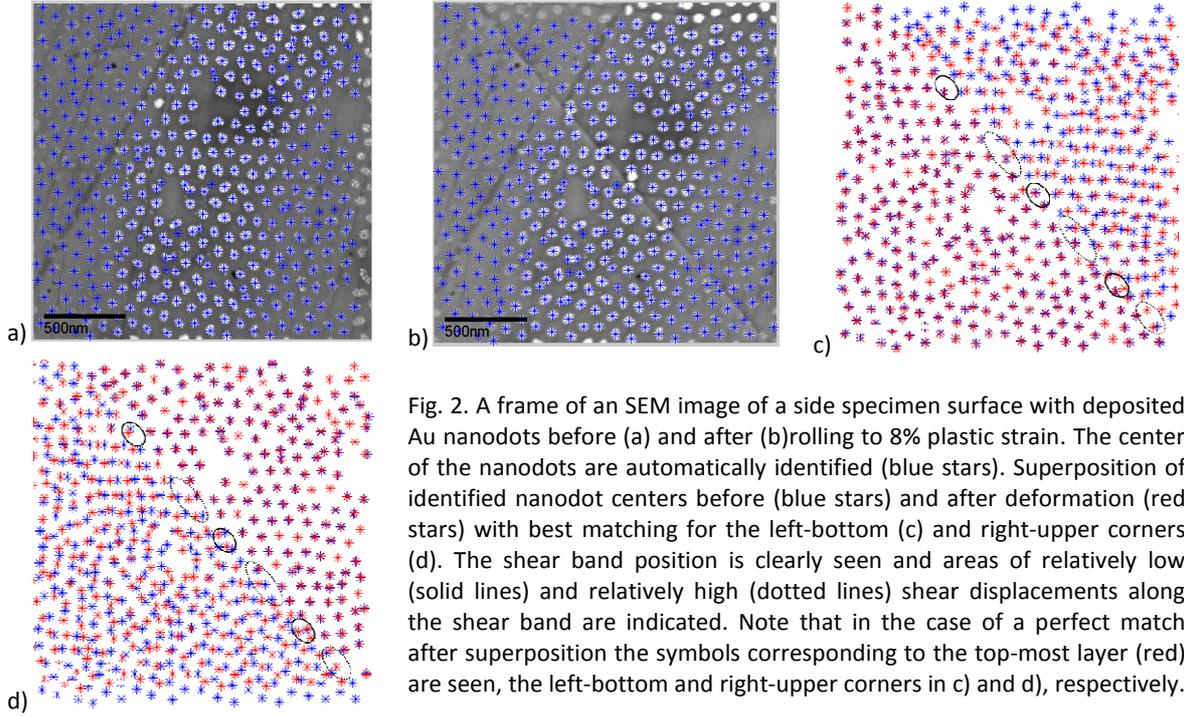

Fig. 2. A frame of an SEM image of a side specimen surface with deposited Au nanodots before (a) and after (b) rolling to 8% plastic strain. The center of the nanodots are automatically identified (blue stars). Superposition of identified nanodot centers before (blue stars) and after deformation (red stars) with best matching for the left-bottom (c) and right-upper corners (d). The shear band position is clearly seen and areas of relatively low (solid lines) and relatively high (dotted lines) shear displacements along the shear band are indicated. Note that in the case of a perfect match after superposition the symbols corresponding to the top-most layer (red) are seen, the left-bottom and right-upper corners in c) and d), respectively.

From the comparison of the center positions of identical nanodots before and after the deformation (marked in Fig. 2a,b), the local displacements $u_{i,j}$ were calculated. Cauchy's infinitesimal strain tensor, $\varepsilon_{ij}$, by definition, is related to the displacement field $u_{i,j}$ by:

$$\varepsilon_{ij} = \frac{1}{2}\left(\frac{\partial u_i}{\partial x_j} + \frac{\partial u_j}{\partial x_i}\right) \quad (1)$$

andthus contains the volumetric and deviatoric strain tensor components. In the current analysis, measurements are taken on the surface of the material. Thus, only the directions of the displacement field (and the corresponding components of the strain tensor) in plane of the patterned surface are included in the analysis.

In Figure 2 the identified centers of nanodots before and after deformation are superimposed in a way that the right-bottom corners of the matrix or the left-upper ones match, Fig. 2c and d, respectively. These images substantiate that deformation is strongly localized at the shear band (within the limit of the inherent resolution of the method) and practically no strain is seen in the undeformed areas of the matrix, i.e. the blue and red stars almost coincide.

A further important feature can be inferred from Fig. 2c,d, namely a strong heterogeneity (and some quasi-periodicity) in the amount of shear strain along the given shear band. After identifying the corresponding nanodots along the shear band before and after rolling, one finds that areas of relatively large displacements (encircled by dotted lines) are alternating with areas of relatively small (almost zero or even in the opposite direction) displacements (depictured by solid lines). These areas are a few hundreds of nanometers long and encompass a number of individual nanodots that in fact rules out possible artifacts. In the following these features are quantitatively discussed.

Applying DIC, the in-plane strain tensor components were calculated from the local displacements. Figure 3 shows exemplarily the volumetric $\varepsilon_{xx}$ (Fig. 3a), the deviatoric $\varepsilon_{xy}$ (Fig. 3b) and the total shear $(\varepsilon_{xy}+\varepsilon_{yx})/2$ (Fig. 3c) components that were determined on the frame shown in Fig. 2. The values of the local strain tensor components were interpolated between the positions of the nanodots to render the spatial distribution more visible. All images of the strain tensor components in Fig. 3 prove that within the limit of the spatial resolution of the method, which is given by the inter-nanodot distance (100 nm in this case), the deformation is localized at the shear

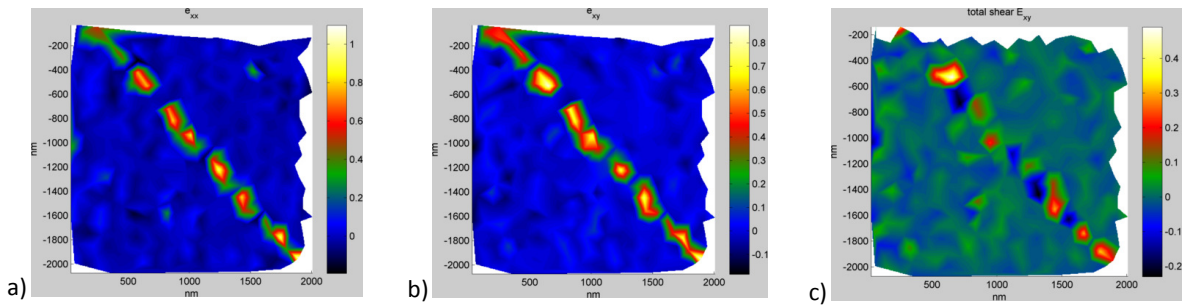

Fig. 3. In-plane components of strain fields associated with the propagation of a shear band (bottom-right to top-left) as determined by DIC: $\varepsilon_{xx}$ (a), $\varepsilon_{xy}$ (b) and total shear $(\varepsilon_{xy}+ \varepsilon_{yx})/2$ (c)

band. This result confirms the view that the major part of the macroscopic strain is accommodated by slip along the shear bands and that the regions that have distances in excess of about 100 nm to the shear bands are not affected by the deformation in terms of plastic yielding (at least within the resolution of the applied method). These results are in variance to earlier reports on the spatial distribution of the indentation modulus [28].

The results obtained by DIC also indicate clearly that adjacent local regions of the shear band show opposite signs for the respective strain tensor components, confirming the direct observation shown in Fig. 2 and indicating that stick-slip like motion along the shear band was occurring during the shear band activation. On average, the total shear strain alternates between +0.30 and -0.10 with maximum values approaching +0.44 and -0.20, Fig. 3c. The length of these regions along the shear band are of the order of 100– 400 nm, in excellent agreement with the length scale of the density alterations inside the shear bands observed by correlative analytical transmission electron microscopy [6,7]. One aspect concerning the results obtained by transmission electron microscopy [6] should be noted here: deviations of opposite sign from the main propagation direction (within several degrees) were observed to accompany the changes of the density of the shear band. Also on larger length scales, shear bands show frequent deviations from the main propagation directions. Such deviations from the main propagation direction are equivalent to local topological maxima and minima along the shear band's path of propagation. Thus, during shear band activation when an external shear stress is applied, the macroscopically uniform slip creates locally regions that are compressed (just before they slip over the nearest topological maximum along the propagation path) and adjacent regions that are dilated, since they just crossed a topological maximum. This picture is in line with the observed distribution of the strain tensor components along the shear band and also with the observation of alternating density changes of opposite sign with similar wavelengths of the order of a few 100 nm, thus explaining that stick-slip motion is also occurring for localized regions of an individual shear band and can also occur under the conditions of rolling. These observations might also indicate that regions of enhanced shear susceptibility are distributed inside the glass on that length scale.

In summary, the current analyses using digital image correlation indicates that the major part of the plastic strain is accommodated in the narrow shear band regions, not exceeding about 100 nm in width. The observation of adjacent regions with alternating and opposite changes of the strain tensor components indicate the occurrence of local stick-slip behavior of regions of individual shear bands with a characteristic length scale on the order of a few 100 nm, in agreement with alternating and opposite density changes of shear band sections observed by correlative analytical transmission electron microscopy. Together with the observed alternating and opposite small deviations of the local shear band orientation with respect to the main propagation direction, these results indicate that the local topology of shear bands, consisting of topological minima and maxima that need to be overcome, result in stick-slip like motion of the material along the activated shear bands thus yielding the observed alterations of density and strain tensor components.

The authors gratefully acknowledge funding by the Deutsche Forschungsgemeinschaft in the framework of SPP 1594.